\documentclass[sigconf]{acmart}

\usepackage[T1]{fontenc}
\usepackage[utf8]{inputenc}
\usepackage{booktabs}
\usepackage{fixltx2e}
\usepackage{multicol}
\usepackage{graphicx}
\usepackage{caption}
\usepackage{amsmath} 
\usepackage{balance}
\usepackage{pifont}
\usepackage{multirow}
\usepackage[colorinlistoftodos,prependcaption,textsize=tiny]{todonotes}
\newcommand{\cmark}{\ding{51}}%
\newcommand{\xmark}{\ding{55}}%

\def\BibTeX{{\rm B\kern-.05em{\sc i\kern-.025em b}\kern-.08emT\kern-.1667em\lower.7ex\hbox{E}\kern-.125emX}}

\copyrightyear{2020}
\acmYear{2020}
\setcopyright{acmlicensed}
\acmConference[CHIIR '20]{2020 Conference on Human Information Interaction and Retrieval}{March 14--18, 2020}{Vancouver, BC, Canada}
\acmBooktitle{2020 Conference on Human Information Interaction and Retrieval (CHIIR '20), March 14--18, 2020, Vancouver, BC, Canada}
\acmPrice{15.00}
\acmDOI{10.1145/3343413.3378009}
\acmISBN{978-1-4503-6892-6/20/03}

\begin{document}
\fancyhead{}
\title{Quantifying the Effects of Prosody Modulation on User Engagement and Satisfaction in Conversational Systems}

\author{Jason Ingyu Choi}
\affiliation{%
  \institution{Computer Science Department\\ Emory University}
%  \city{Atlanta}
%  \state{Georgia}
}
\email{in.gyu.choi@emory.edu }

\author{Eugene Agichtein}
\affiliation{%
  \institution{Computer Science Department\\ Emory University}
%  \city{Atlanta}
%  \state{Georgia}
}
\email{eugene.agichtein@emory.edu}
\renewcommand{\shortauthors}{J. Choi et al.}

\begin{abstract}
As voice-based assistants such as Alexa, Siri, and Google Assistant become ubiquitous, users increasingly expect to maintain natural and informative conversations with such systems. However, for an open-domain conversational system to be coherent and engaging, it must be able to maintain the user's interest for extended periods, without sounding ``boring'' or ``annoying''. In this paper, we investigate one natural approach to this problem, of {\em modulating response prosody}, i.e., changing the pitch and cadence of the response to indicate delight, sadness or other common emotions, as well as using pre-recorded {\em interjections}. Intuitively, this approach should improve the naturalness of the conversation, but attempts to {\em quantify} the effects of prosodic modulation on user satisfaction and engagement remain challenging. To accomplish this, we report results obtained from a large-scale empirical study that measures the effects of prosodic modulation on user behavior and engagement across multiple conversation domains, both immediately after each turn, and at the overall conversation level. Our results indicate that the prosody modulation significantly increases both immediate and overall user satisfaction. However, since the effects vary across different domains, we verify that prosody modulations do not substitute for coherent, informative content of the responses. Together, our results provide useful tools and insights for improving the naturalness of responses in conversational systems.
\end{abstract}

\maketitle

\section{Introduction and Background}
With the proliferation of voice-based assistants such as Alexa, Siri, and Google Assistant, there has been a resurgence of research into building truly intelligent conversational assistants that can maintain a long, natural conversation with users. One important direction was a recent series of Amazon Alexa Prize Challenges, providing a competition platform and monetary incentives to spur development \citep{alexaprize1, alexaprize2} in conversational AI. Many practical applications of conversational systems have been proposed (e.g., for companionship to improve mental well-being \citep{morris2018towards}, and for therapy \citep{fitzpatrick2017delivering}). While much room for improvement remains in the current implementations of the conversational AI systems, the potential for intelligent, empathetic and broad-coverage conversational systems is widely recognized.

\begin{figure}[t!]
\includegraphics[height=200pt]{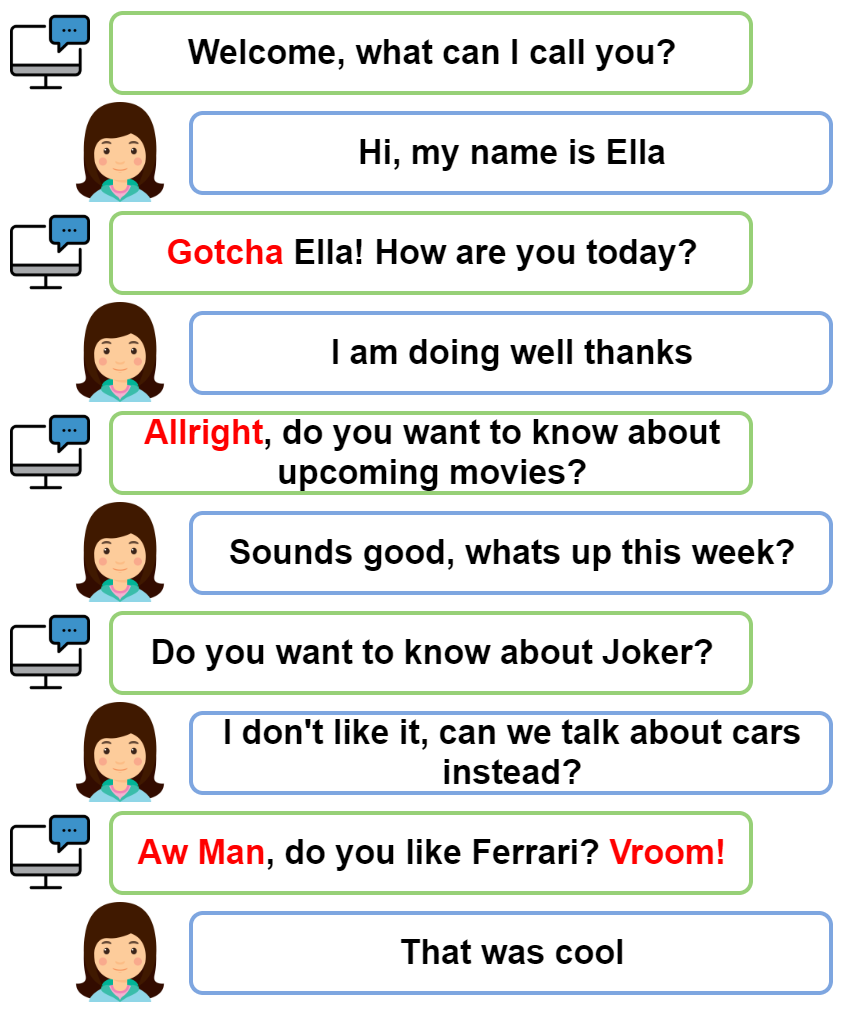}
\vspace{-0.2cm}
\captionof{figure}{Sample human-machine conversation from our system. The red texts show response examples after inserting prerecorded Speechcons to convey artificial emotion.}
\vspace{-0.2cm}
\label{sample_conv}
\end{figure}

However, for open-domain conversational system to be engaging and intelligent, it must keep the user's interest for extended periods, without sounding ``boring'' or ``annoying'', which unfortunately is the case for current voice-based assistants. In this paper, we investigate one natural approach to handle ``boring'' responses, which is to {\em modulate response prosody} via commonly available Speech Synthesis Markup Language (SSML) \citep{SSML}. For our experiments, we replaced common phrases (i.e. filter words or interjections) with prerecorded Speechcons from Alexa Skills Kit APIs\footnote{\url{https://https://developer.amazon.com/en-US/docs/alexa/custom-skills/speechcon-reference-interjections-english-us.html}}. In some cases, the pitch and rate of these Speechcons are additioanlly tuned to convey excitement, hesitation and emphasis, allowing the agent to deliver a variety of empathetic responses to users. The example conversation\footnote{Due to the Alexa Prize data confidentiality rules, we cannot reproduce an actual user conversation, but the example represents a typical conversation with our system.} provided in Figure \ref{sample_conv} shows how our system utilized prerecorded Speechcons such as ``Allright'' or ``Aw Man'' to improve naturalness in conversations. 

%These systems do not naturally interact with the users or express emotion, whether they are responding positively, negatively, or reporting failure.

As our main contribution, we report the results of a large-scale, empirical study aiming to quantify {\em overall} and {\em immediate} effects of prosody modulation on user engagement and satisfaction, in open-domain, unrestricted conversational setting as part of the Amazon Alexa Prize competition. Our results indicate that the prosody modulation significantly increases both types of user satisfaction, but the degree of improvements varies across different domains. Specifically, our contributions include:

\begin{itemize}
\item One of the first attempt to quantify the immediate effects of prosodic modulation on user satisfaction and engagement in unrestricted open-domain conversations
\item A large-scale empirical experiment comparing the effects of prosody modulation to user satisfaction across multiple conversation domains
\end{itemize}

Next, we briefly review closely related work to place our contributions in context.

\subsection{Background and Related Work}
\label{sec:related}

\paragraph{\bf Conversational AI and Satisfaction Prediction}
Recently, conversational systems research has experienced dramatic progress. For example, automatic speech recognition (ASR) has been revolutionized by neural models \citep{speech_RNN}. Similarly for dialogue management, both rule-based \citep{form-DM, ravenclaw} and end-to-end \citep{DM_end, luo2018learning} systems were studied extensively. To maintain a flexible and scalable structure, several architectures have been proposed \citep{gunrock, sounding_board} as well. 

As these systems became more sophisticated, many work proposed new ideas to automate the evaluation process by predicting conversational user satisfaction, as defined in \citep{user_satisfaction_theory1, paradise1, paradise2}. For instance, there have been successful attempts to predict satisfaction once conversations (sessions) are completed, using traditional methods \citep{kiseleva2016predicting, egregious} and neural-based models \citep{hashemi2018measuring, hancock2019learning}. Lastly, one recent work \citep{ConvSAT} proposed a unified neural framework to predict offline (session-level) and online (turn-level) satisfaction simultaneously. 

\paragraph{\bf Speech Synthesis and Prosody Modulations}
Speech synthesis is an active research field that studies the artificial production of human speech \citep{tts_overview}. Speech synthesizers, also known as text-to-speech (TTS) synthesizers, are placed at the final phase of modern dialogue systems to transform textual output into a natural voice output. In conversation-related speech strategies, several work focused on analyzing the impact of interjections and filter words such as ``Um'', ``Uh'' and ``Wow'' to user behaviors. For instance, \citep{um1, um2} reported the change in eye gaze behavior when conversations start with these filter words. Another study \citep{syrdal2010speech} showed that speech models trained on positive exclamations achieved higher satisfaction from listening tests. Based on these empirical findings, recent dialogue systems \citep{sounding_board, gunrock, bowden2019slugbot, ji2017two} incorporated liveliness by adjusting the prosodies of interjections and filter words using SSML.

However, studies to evaluate the impact of prosody modifications to user satisfaction have been limited. One recent study \citep{chuklin2019using} measured the effectiveness of prosody modification using crowd-sourced workers and showed that while comprehensiveness (i.e. informative, correctness) improved, naturalness (i.e. interruption) decreased. However, this evaluation was measured only on each information-seeking turn and the authors highlighted the need for future study in a more realistic conversation. Cohn et al. \citep{gunrock_ssml} addressed this limitation and showed modifying both filter words and interjections achieved the highest user ratings when evaluated on large-scale open-domain conversations. However, since user ratings convey an overall impression, quantifying the immediate effects of prosody modification within each conversation remains unexplored.

Thus, our work extends the ideas here by first train a state of the art immediate- and offline- satisfaction prediction model \citep{ConvSAT} and quantify both immediate and longer-term effects on user satisfaction and engagement using our proposed metrics, which are described later.

\section{Experimental Setup}
In this section, we present our conversational system and data collection process, followed by a setup to train an online satisfaction model. Lastly, several evaluation metrics are proposed.

\subsection{Conversational System and Dataset}
\label{sec-system}

\paragraph{\bf Alexa Prize 2018}
This study was performed as part of a naturalistic assessment of open-domain conversational systems, organized by the Amazon Alexa Prize Conversational AI challenge. Amazon Alexa customers were randomly assigned to each participating system, and could converse on a wide range of topics. At the end of the conversation, the customer could optionally leave a rating (1.0-5.0) and optional comment feedback. It is worth emphasizing that one of the main goals of the competition was to design an agent capable of maintaining an engaging conversation with a user for 20 minutes, which required significant engineering effort, outlined below, to enable the collection of informative and realistic conversational data.

\paragraph{\bf System Architecture}
Our goal was to develop a conversational agent that helps the user be informed about the world around them, while being entertained and engaged. Our agent incorporated real-time search, informed advice, and the latest information into the conversation by attempting to discuss and share information on many popular domains. To do so, our system had to accurately detect the user’s intent from the combinations of explicitly stated and implied evidence from the context. The detailed description of the agent architecture, dialogue management, response ranking and generation is reported in reference \citep{emory}. We provide brief descriptions of some of the most popular domain-specific components (or mini-agents), which we selected to investigate the benefits and effects of modulating the prosody of the agent's responses. 

% \begin{figure}[h]
% \includegraphics[height=225pt]{irisbot.png}
% \captionof{figure}{Sample human-machine conversation, enhanced with speech synthesis markup language (SSML).}
% \label{sample_conv}
% \end{figure}

% \paragraph{\bf Supported Popular Domains}
\begin{itemize}
\item{\textit{Opening}}: Introduction begins with a required greeting to identify the agent as a specialized Alexa skill, and attempts to ``break the ice'' with the user by exchanging names, and proposing initial topics for discussion. 

% This component establishes the initial tone of the whole conversation, and we conjectured that it has to sound as natural and fluent as possible for the user to continue chatting with the bot about more specialized topics such as Movies or Music.

\item{\textit{Movies}}: Movies component can hold in-depth conversations on most movie-related topics including trending movies, TV shows, actor/director information and personalized movie recommendations.

% To accomplish this, this component uses a variety of question and answer templates (which we call "chat-mode") to provide the framework of a natural conversation, as well as determine customer preferences to deal with the initial "cold-start" problem. However, we found that to keep the customer engaged, we must also share immediately useful or entertaining information instead of just asking questions.

\item{\textit{Music}}: Music component handles popular music-related questions such as trending chart by genre, upcoming concert information and music recommendations. 

% Just as Movies component does, the Music component also attempts to chat with the customer to gather indicators of interest for music genres or sub-genres, artists, and attempts to offer similar artists, genres, or provide general information of interest.

\item{\textit{News}}: News component is responsible for updating the customer with trending news or news on specific entities. It covers a wide range of popular news domains such as politics, science, celebrity, sports and so on.

% This component aggregates data from multiple news sources, including the provided news articles and summaries, and attempts to answer questions about the news using both local news corpus and publicly accessible search engine APIs.

\item{\textit{Games}}: Games component can chat and recommend the most popular upcoming games for various gaming platforms such as PlayStation, Xbox and PC. 

\item{\textit{Travel}}: Travel component supports real-time place search such as retrieving recent reviews, ratings and addresses.
\end{itemize}

\paragraph{\bf Dataset Overview}
Our main dataset is conversations collected during the 2nd-round Alexa Prize 2018 competition. For this study, we collected logs from two specific versions A and B, which correspond to the system versions before and after adding prosody effects to our system responses. This controlled setup is to eliminate any potential change to different parts of the system that may affect the integrity of this evaluation. Version A and B were live during July 23rd - July 27th and July 25th - July 31st. Please note that the overlap between these two periods is expected because our production server had 8 different instances for traffic control and A/B testing. Table \ref{alexa_stats} summarizes the statistics of two datasets A and B, each obtained from version A and B respectively.

\begin{table}[h]
    \centering
    \begin{tabular}{l|c|c}
        \toprule
        & \textbf{Dataset A} & \textbf{Dataset B} \\
        \toprule
        \textbf{Prosody} & \xmark & \cmark \\
        \textbf{Dialogues} & 1659 & 1202 \\
        \textbf{Rated Dialogues} & 984 (59.3\%) & 670 (55.7\%) \\
        \textbf{Average User Ratings} & 3.43 & 3.47 \\
        \textbf{Average Turns} & 17.51 & 17.29 \\
        \bottomrule
    \end{tabular}
    \caption{Statistics on two datasets A and B,  collected immediately before (A) and after (B) adding prosody modulation.}
    \label{alexa_stats}
    \vspace{-0.2cm}
\end{table}

In general, both datasets have similar statistics. Even though dataset A has a slightly larger number of conversations than dataset B, the difference in averaged number of turns is small. The standard deviations of number of turns distributions are 14.75 and 14.36, indicating the diversity in conversation lengths for both datasets. Dataset A also has a slightly higher fraction of rated dialogues. After adding the prosody effect, there is a small increase of 0.04 in averaged user ratings. We emphasize that the only difference between these two datasets is the presence of prosody modification in system responses.

% Figure \ref{engagement_stats} shows the count of extracted engagements for two datasets, categorized by each domain. Please note that users were allowed to skip openings and start engaging.

% % \vspace{-2mm}
% \begin{table}[h]
%     \centering
%     \begin{tabular}{l|c|c}
%         \toprule
%         & \textbf{Dataset A} & \textbf{Dataset B} \\
%         \toprule
%         \textbf{Opening} & 1514 & 1062 \\
%         \textbf{Movies} & 672 & 377 \\
%         \textbf{Music} & 569 & 328 \\
%         \textbf{News} & 573 & 310 \\
%         \textbf{Travel} & 297 & 19 \\
%         \textbf{Video games} & 378 & 25 \\
%         \textbf{Total} & 4003 & 2121 \\ 
%         \bottomrule
%     \end{tabular}
%     \caption{Count of extracted engagements (depth <= 2) for different domains.}
%     \label{engagement_stats}
% \end{table}
% \vspace{-6mm}

\begin{table*}[t!]
    \centering
    \begin{tabular}{l|l|l|l|l|c}
        \toprule
        \textbf{Domains} & \textbf{\textit{SAT\textsubscript{immediate}}} & \textbf{\textit{SAT\textsubscript{engagement}}} & \textbf{Depth} & \textbf{Samples} & \textbf{Prosody}  \\
        \toprule
        \textbf{Opening} & 0.530 & 1.644 & 2.812 & 1514 & \multirow{7}{*}{\xmark} \\
        \textbf{Movies} & 0.443 & 2.111 & 3.631 & 672 &  \\
        \textbf{Music} & \textbf{0.454} (-8.8\%) & 1.535 & 3.506 & 569 &  \\
        \textbf{Games} & 0.380 & 1.685 & 3.666 & 573 &  \\
        \textbf{Travel} & 0.443 & 1.563 & 3.427 & 297 &  \\
        \textbf{News} & 0.413 & 1.274 & 3.555 & 378 &  \\
        \textbf{All} & 0.457 & 1.672 & 3.289 & 4003 &  \\
        \toprule
        \textbf{User ratings} & \multicolumn{5}{c}{3.43} \\
        \toprule

        \textbf{Opening} & \textbf{0.536} (+1.1\%) & \textbf{1.705} (+3.7\%)\textbf{*} & \textbf{3.00} (+6.7\%)\textbf{*} & 1062 & \multirow{7}{*}{\cmark} \\
        \textbf{Movies} & \textbf{0.576} (+30.0\%)\textbf{*} & \textbf{2.137} (+2.1\%) & \textbf{3.790} (+4.4\%) & 377 &  \\
        \textbf{Music} & 0.414 & \textbf{1.656} (+7.8\%)\textbf{*} & \textbf{3.670} (+4.8\%) & 328 &  \\
        \textbf{Games} & \textbf{0.499} (+31.3\%)\textbf{*} & \textbf{1.718} (+1.9\%) & \textbf{3.790} (+3.3\%) & 310 &  \\
        \textbf{Travel} & \textbf{0.738} (+66.5\%)\textbf{*} & \textbf{2.047} (+30.9\%)\textbf{*} & \textbf{4.578} (+32.1\%)\textbf{*} & 19 &  \\
        \textbf{News} & \textbf{0.426} (+3.1\%) & \textbf{1.624} (+27.4\%)\textbf{*} & \textbf{4.800} (+35.0\%)\textbf{*} & 25 &  \\
        \textbf{All} & \textbf{0.516} (+12.9\%)\textbf{*} & \textbf{1.778} (+6.3\%)\textbf{*} & \textbf{3.395} (+3.2\%)\textbf{*}& 2121 &  \\
        \toprule
        \textbf{User ratings} & \multicolumn{5}{c}{\textbf{3.47} (+1.1\%)} \\
        \bottomrule
    \end{tabular}
    \vspace{2mm}
    \caption{Change in online satisfaction difference (\textit{SAT\textsubscript{immediate}}), engagement-level satisfaction difference (\textit{SAT\textsubscript{engagement}}), conversation depth and averaged user ratings before (\xmark) and after (\cmark) adding prosody modification. ``*'' indicates statistical significance of improvement based on two-tailed Student's t-test with $p<0.05$.}
    \label{main_results}
\vspace{-5mm}
\end{table*}

\subsection{Evaluation Setup}
\paragraph{\bf Training Online Satisfaction Model}
To obtain ground-truth data for online satisfaction prediction, we randomly sampled 100 conversations and recruited two human annotators to label satisfaction $\in$ [1.0, 2.0, 3.0, 4.0, 5.0] for each turn by only considering the past information so far. Since our dataset is private, both annotators were internally recruited from our team (not crowd-sourced). The agreement between two annotators was substantial according to Kappa score of \textbf{0.753} \citep{kappa}. In case of disagreements, two labels were averaged to minimize disagreement. 

For training data, we followed the identical data programming rule proposed in \citep{ConvSAT}, which defines sets of heuristic labeling functions using system states and user behavioral signals to generate weak labels. This setup is required because human annotation is time-consuming and authors highlighted the effectiveness in data programming to generate large-scale training data from unlabeled conversations.

Using the generated training data and manually annotated test data, a LSTM-based online satisfaction prediction model \citep{ConvSAT} was trained. Because our goal is to measure exact changes in satisfaction across different turns, we trained the model in a regression setting to minimize the mean squared loss between predicted and annotated ratings. Thus, we emphasize that even though the training labels were discrete, the model was trained to predict a continuous range of ratings. We believe regression fits better to quantify the change in user satisfaction than predicting discrete labels (i.e. counting not satisfied vs. satisfied). Initially, heuristically generated labels scored \textbf{1.243} mean absolute error (MAE) on the test set. After training, the model achieved \textbf{0.772} MAE on the test set. Using this pre-trained model, all the turns in the two datasets are annotated with predicted satisfaction values.

\paragraph{\bf Evaluation Metrics}
We define engagements within conversations as sub-conversations that have 2 or more depth within the same domains. Engagements are extracted from 6 different domains defined in Subsection \ref{sec-system}. For instance, the conversation illustrated in Figure \ref{sample_conv} has three distinct engagements, which are \textit{opening} (depth=2), \textit{movies} (depth=2) and \textit{cars} (ongoing). These domains are selected because they were the most popular, but most importantly, the earliest domains to utilize prosody modifications. Since other domains incorporated prosody modifications after version B, they were excluded from this study. We propose metrics in four different dimensions to measure user satisfaction (\textit{SAT}): 1) immediate online satisfaction; 2) engagement-level satisfaction; 3) engagement depth; 4) user ratings.

First, we propose to capture the immediate effect on the predicted satisfaction after responses with  prosody modifications, by computing the changes in the immediate satisfaction for the current turn (SAT\textsubscript{i}) and the next turn (SAT\textsubscript{i+1}). This is equivalent to measuring the difference in predicted satisfaction before and after the prosody modulation. These differences are summed and normalized by the count (\textit{N}) of (SAT\textsubscript{i}, SAT\textsubscript{i+1}) pair per domain. We compute this metric as an immediate satisfaction difference (\textit{SAT\textsubscript{immediate}}):
\begin{gather}
    \textit{SAT\textsubscript{immediate}} = \frac{\sum_{1}^{N} (SAT\textsubscript{i+1}-SAT\textsubscript{i})}{N}
\end{gather}

We also compute the engagement-level difference in satisfaction (\textit{SAT\textsubscript{engagement}}) from the starting (SAT\textsubscript{i}) and ending (SAT\textsubscript{i+depth}) satisfaction of each engagement, with same normalization scheme where \textit{N} is the total count of engagements per domain:
%\vspace{-3mm}
\begin{gather}
    \textit{SAT\textsubscript{engagement}} = \frac{\sum_{1}^{N} (SAT\textsubscript{i+depth}-SAT\textsubscript{i})}{N}
\end{gather}

Finally, we measure the differences in engagement {\em depth}, that is, the average number of turns a user spends conversing with each component. These three metrics are first computed on domain-specific level, and aggregated to measure the overall effect. Lastly, we report the averaged user satisfaction ratings (self-reported by Alexa users) to highlight the overall impact. 
\label{evaluation_metrics}

%\vspace{-0.2cm}
\section{Results and Discussion}
In this section, we report the effect of prosody modulation on user satisfaction and engagement on six different domains.%, followed by a conclusion.

%\vspace{-0.2cm}
\paragraph{\bf Evaluation of Prosody Modulation}
According to the results reported in Table \ref{main_results}, the results are promising as they show improvements in all three metrics on diverse domains. When the results are aggregated for all six domains, there are 12.9\%, 6.3\% and 3.2\% improvement on \textit{SAT\textsubscript{immediate}},  \textit{SAT\textsubscript{engagement}} and depth, respectively. These improvements are statistically significant based on two-tailed Student's t-test (unpaired) with $p<0.05$. The slight increase in user ratings from 3.43 to 3.47 further confirms that the improvements reflect the increased perceived quality in conversations.

Openings started with prosody modifications show improvements in all metrics compared to the openings without prosody modifications. 1.1\% increase in openings (\textit{SAT\textsubscript{immediate}}) is particularly interesting because we are measuring the change that is not conditioned to any previous context. We claim that the initial prosody modifications create a more positive first impression of our system, subsequently increasing \textit{SAT\textsubscript{engagement}} and decreasing the likelihood to skip openings.

For each domain, \textit{Travel} showed the strongest improvements on \textit{SAT\textsubscript{immediate}} and \textit{SAT\textsubscript{engagement}} metrics while \textit{News} achieved the most increase in depth with statistical significance. One limitation is that the samples on these two domains are much less compared to other domains. \textit{Movies} and \textit{Games} domain, when evaluated on hundreds of samples, show that there are 30.0\% and 31.3\% statistically significant improvements on \textit{SAT\textsubscript{immediate}}. Depth and \textit{SAT\textsubscript{engagement}} increased as well, but the changes are not statistically significant.  

Surprisingly, for \textit{Music} domain, there is a decrease in immediate satisfaction after prosody modifications. Unlike the \textit{Travel} and \textit{Games} components, where modified interjections occurred multiple times between engagements, \textit{Music} conversations only modulated prosody rarely and not in a consistent way, indicating that prosody modulation must be carefully matched to the target domain, as we plan to explore in future work. In summary, our results showed that while overall both engagement and satisfaction increased when an agent becomes less monotonous and more ``natural'', the benefits vary across domains. For \textit{Games}, \textit{News}, and \textit{Travel} domains the improvements are particularly noticeable, and less so for \textit {Music} and \textit{Movies} domains. %Our results indicate that informativeness and coherence of the responses remain critical and cannot be masked by natural-sounding responses. 
%\vspace{-0.2cm}

\section{Conclusions and Future Work}
We reported results of a large scale, real-world evaluation of the effects of modulating prosody for conversational agent responses in several dimensions. Specifically, we confirmed that prosody modulation significantly effects immediate user satisfaction with an agent's responses, and that in some cases can also significantly increase the engagement of the users with the system, ultimately improving the overall subjective self-reported satisfaction ratings. Our analysis was based on thousands of conversations of real customers with an open-domain conversational agent, deployed as part of the Amazon Alexa Prize 2018 competition. While the overall improvements were significant, the effects were more dramatic in some domains, such as \textit{Games} and \textit{Travel}.

Despite these promising results, further study is needed. We conjecture that in addition to the domain-based differences in appropriate prosody modulation, personal differences between users may also effect the appropriate prosody for responses. In future work, we plan to investigate the effects of incorporating automatically learned prosody modulation techniques, e.g.~\cite{skerry2018tacotron}, potentially fine-tuned for each domain, in order to generate more contextually relevant and natural-sounding responses. 

%\vspace{-0.2cm}
\paragraph{\bf Acknowledgements}
We gratefully acknowledge financial, technical, and computational support from Amazon Alexa Prize 2018.

% \begin{acks}
% We gratefully acknowledge the computational and technical support from Amazon Alexa Prize 2017 and 2018.
% \end{acks}

\newpage
\balance
\bibliographystyle{abbrv}
\bibliography{acmart}

\begin{thebibliography}{10}

\bibitem{emory}
A.~Ahmadvand, I.~Choi, H.~Sahijwani, J.~Schmidt, M.~Sun, S.~Volokhin, Z.~Wang,
  and E.~Agichtein.
\newblock Emory irisbot: An open-domain conversational bot for personalized
  information access.
\newblock {\em Proc. Alexa Prize}, 2018.

\bibitem{ravenclaw}
D.~Bohus and A.~I. Rudnicky.
\newblock Ravenclaw: Dialog management using hierarchical task decomposition
  and an expectation agenda.
\newblock In {\em Eighth European Conference on Speech Communication and
  Technology}, 2003.

\bibitem{bowden2019slugbot}
K.~K. Bowden, J.~Wu, W.~Cui, J.~Juraska, V.~Harrison, B.~Schwarzmann,
  N.~Santer, and M.~Walker.
\newblock Slugbot: Developing a computational model andframework of a novel
  dialogue genre.
\newblock {\em arXiv preprint arXiv:1907.10658}, 2019.

\bibitem{gunrock}
C.-Y. Chen, D.~Yu, W.~Wen, Y.~M. Yang, J.~Zhang, M.~Zhou, K.~Jesse, A.~Chau,
  A.~Bhowmick, S.~Iyer, et~al.
\newblock Gunrock: Building a human-like social bot by leveraging large scale
  real user data.
\newblock {\em Alexa Prize Proceedings}, 2018.

\bibitem{ConvSAT}
I.~J. Choi, A.~Ahmadvand, and E.~Agichtein.
\newblock Offline and online satisfaction prediction in open-domain
  conversational systems.
\newblock In {\em Proceedings of the 28th ACM International Conference on
  Information and Knowledge Management}. ACM, 2019.

\bibitem{chuklin2019using}
A.~Chuklin, A.~Severyn, J.~R. Trippas, E.~Alfonseca, H.~Silen, and D.~Spina.
\newblock Using audio transformations to improve comprehension in voice
  question answering.
\newblock In {\em International Conference of the Cross-Language Evaluation
  Forum for European Languages}, pages 164--170. Springer, 2019.

\bibitem{gunrock_ssml}
M.~Cohn, C.-Y. Chen, and Z.~Yu.
\newblock A large-scale user study of an alexa prize chatbot: Effect of tts
  dynamism on perceived quality of social dialog.
\newblock In {\em Proceedings of the 20th SIGdial Workshop on Discourse and
  Dialogue, Cited by}, volume~2, 2019.

\bibitem{tts_overview}
T.~Dutoit.
\newblock High-quality text-to-speech synthesis: An overview.
\newblock {\em Journal Of Electrical And Electronics Engineering Australia},
  17:25--36, 1997.

\bibitem{sounding_board}
H.~Fang, H.~Cheng, E.~Clark, A.~Holtzman, M.~Sap, M.~Ostendorf, Y.~Choi, and
  N.~A. Smith.
\newblock Sounding board--university of washington’s alexa prize submission.
\newblock {\em Alexa Prize Proceedings}, 2017.

\bibitem{fitzpatrick2017delivering}
K.~K. Fitzpatrick, A.~Darcy, and M.~Vierhile.
\newblock Delivering cognitive behavior therapy to young adults with symptoms
  of depression and anxiety using a fully automated conversational agent
  (woebot): a randomized controlled trial.
\newblock {\em JMIR mental health}, 4(2):e19, 2017.

\bibitem{form-DM}
D.~Goddeau, H.~Meng, J.~Polifroni, S.~Seneff, and S.~Busayapongchai.
\newblock A form-based dialogue manager for spoken language applications.
\newblock In {\em Spoken Language, 1996. ICSLP 96. Proceedings., Fourth
  International Conference on}, volume~2, pages 701--704. IEEE, 1996.

\bibitem{speech_RNN}
A.~Graves, A.-r. Mohamed, and G.~Hinton.
\newblock Speech recognition with deep recurrent neural networks.
\newblock In {\em Acoustics, speech and signal processing (icassp), 2013 ieee
  international conference on}, pages 6645--6649. IEEE, 2013.

\bibitem{hancock2019learning}
B.~Hancock, A.~Bordes, P.-E. Mazare, and J.~Weston.
\newblock Learning from dialogue after deployment: Feed yourself, chatbot!
\newblock {\em arXiv preprint arXiv:1901.05415}, 2019.

\bibitem{hashemi2018measuring}
S.~H. Hashemi, K.~Williams, A.~El~Kholy, I.~Zitouni, and P.~A. Crook.
\newblock Measuring user satisfaction on smart speaker intelligent assistants
  using intent sensitive query embeddings.
\newblock In {\em Proceedings of the 27th ACM International Conference on
  Information and Knowledge Management}, pages 1183--1192. ACM, 2018.

\bibitem{ji2017two}
J.~Ji, Q.~Wang, Z.~Battad, J.~Gou, J.~Zhou, R.~Divekar, C.~Carlson, and M.~Si.
\newblock A two-layer dialogue framework for authoring social bots.
\newblock {\em Alexa Prize Proceedings. Seattle, WA: Amazon}, 2017.

\bibitem{alexaprize2}
C.~Khatri, B.~Hedayatnia, A.~Venkatesh, J.~Nunn, Y.~Pan, Q.~Liu, H.~Song,
  A.~Gottardi, S.~Kwatra, S.~Pancholi, et~al.
\newblock Advancing the state of the art in open domain dialog systems through
  the alexa prize.
\newblock {\em arXiv preprint arXiv:1812.10757}, 2018.

\bibitem{kiseleva2016predicting}
J.~Kiseleva, K.~Williams, A.~Hassan~Awadallah, A.~C. Crook, I.~Zitouni, and
  T.~Anastasakos.
\newblock Predicting user satisfaction with intelligent assistants.
\newblock In {\em Proceedings of the 39th International ACM SIGIR conference on
  Research and Development in Information Retrieval}, pages 45--54. ACM, 2016.

\bibitem{luo2018learning}
L.~Luo, W.~Huang, Q.~Zeng, Z.~Nie, and X.~Sun.
\newblock Learning personalized end-to-end goal-oriented dialog.
\newblock In {\em Proceedings of the AAAI Conference on Artificial
  Intelligence}, volume~33, pages 6794--6801, 2019.

\bibitem{morris2018towards}
R.~R. Morris, K.~Kouddous, R.~Kshirsagar, and S.~M. Schueller.
\newblock Towards an artificially empathic conversational agent for mental
  health applications: system design and user perceptions.
\newblock {\em Journal of medical Internet research}, 20(6):e10148, 2018.

\bibitem{um2}
L.~M. Pfeifer and T.~Bickmore.
\newblock Should agents speak like, um, humans? the use of conversational
  fillers by virtual agents.
\newblock In {\em International Workshop on Intelligent Virtual Agents}, pages
  460--466. Springer, 2009.

\bibitem{alexaprize1}
A.~Ram, R.~Prasad, C.~Khatri, A.~Venkatesh, R.~Gabriel, Q.~Liu, J.~Nunn,
  B.~Hedayatnia, M.~Cheng, A.~Nagar, et~al.
\newblock Conversational ai: The science behind the alexa prize.
\newblock {\em arXiv preprint arXiv:1801.03604}, 2018.

\bibitem{egregious}
T.~Sandbank, M.~Shmueli-Scheuer, J.~Herzig, D.~Konopnicki, J.~Richards, and
  D.~Piorkowski.
\newblock Detecting egregious conversations between customers and virtual
  agents.
\newblock {\em arXiv preprint arXiv:1711.05780}, 2017.

\bibitem{skerry2018tacotron}
R.~Skerry-Ryan, E.~Battenberg, Y.~Xiao, Y.~Wang, D.~Stanton, J.~Shor, R.~Weiss,
  R.~Clark, and R.~A. Saurous.
\newblock Towards end-to-end prosody transfer for expressive speech synthesis
  with tacotron.
\newblock In {\em International Conference on Machine Learning}, pages
  4693--4702, 2018.

\bibitem{syrdal2010speech}
A.~K. Syrdal, A.~Conkie, Y.-J. Kim, and M.~C. Beutnagel.
\newblock Speech acts and dialog tts.
\newblock In {\em Seventh ISCA Workshop on Speech Synthesis}, 2010.

\bibitem{SSML}
P.~Taylor and A.~Isard.
\newblock Ssml: A speech synthesis markup language.
\newblock {\em Speech communication}, 21(1-2):123--133, 1997.

\bibitem{um1}
J.~E.~F. Tree.
\newblock Listeners' uses ofum anduh in speech comprehension.
\newblock {\em Memory \& cognition}, 29(2):320--326, 2001.

\bibitem{kappa}
A.~J. Viera, J.~M. Garrett, et~al.
\newblock Understanding interobserver agreement: the kappa statistic.
\newblock {\em Fam Med}, 37(5):360--363, 2005.

\bibitem{paradise1}
M.~A. Walker, D.~J. Litman, C.~A. Kamm, and A.~Abella.
\newblock Paradise: A framework for evaluating spoken dialogue agents.
\newblock In {\em Proc. of ACL}, pages 271--280. Association for Computational
  Linguistics, 1997.

\bibitem{paradise2}
M.~A. Walker, R.~Passonneau, and J.~E. Boland.
\newblock Quantitative and qualitative evaluation of darpa communicator spoken
  dialogue systems.
\newblock In {\em Proc. of ACL}, pages 515--522. Association for Computational
  Linguistics, 2001.

\bibitem{user_satisfaction_theory1}
B.~H. Wixom and P.~A. Todd.
\newblock A theoretical integration of user satisfaction and technology
  acceptance.
\newblock {\em Information systems research}, 16(1):85--102, 2005.

\bibitem{DM_end}
X.~Yang, Y.-N. Chen, D.~Hakkani-T{\"u}r, P.~Crook, X.~Li, J.~Gao, and L.~Deng.
\newblock End-to-end joint learning of natural language understanding and
  dialogue manager.
\newblock In {\em Acoustics, Speech and Signal Processing (ICASSP), 2017 IEEE
  International Conference on}, pages 5690--5694. IEEE, 2017.

\end{thebibliography}
\end{document}